\documentclass[12pt]{article}
\usepackage{graphicx}
\usepackage{subfigure}
\usepackage{amssymb}

\begin{document}
\title{Varying G dynamics}
\author{B.G.Sidharth \footnote{Email:iiamisbgs@yahoo.co.in}\\International Institute of Applicable Mathematics\\ and
Information Sciences\\
Adarsh Nagar, Hyderabad \\ \vspace{.3cm} \& \\  B.S.Lakshmi
\footnote{Email:bslakshmi2000@yahoo.com}\\ Department of Mathematics, JNTUCEH\\
Kukatpally, Hyderabad}

\date{}
\maketitle

\begin{abstract}
In this paper it is shown that dynamics based on a
variation of the gravitational constant $G$ with time solves several
puzzling and anomalous features observed, for example the rotation
curves of galaxies (attributed to as yet undetected Dark matter). It
is also pointed out that this provides an explanation for the
anomalous acceleration of the Pioneer space crafts observed by
J.D.Anderson and co-workers.

\end{abstract}

\section{Introduction}

The Milky Way to which our sun belongs, contains about 100 billion
stars. On even larger scales, individual galaxies are concentrated
into clusters of galaxies. These clusters consist of the galaxies
and any material which is
 in the space between the galaxies. The force that holds the cluster together is gravity. The space between galaxies in clusters
 is filled with a hot gas.
The cluster includes the galaxies and any material which is in the
space between the galaxies. The gas is hot enough to make this space
shine in X-rays instead of visible light. By studying the
distribution and temperature of the hot gas we can measure the force
of gravity from all the material in the cluster. This allows us to
determine  the total material content in that cluster.

\begin{figure}[htp]  \label{fig:1}
  \begin{center}
    \subfigure[Expected Keplerian
    curves]{\label{fig:1a}\includegraphics[scale=.65]{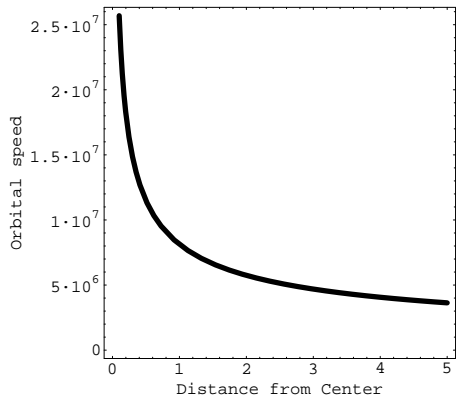}}
    \quad \quad
    \subfigure[Observed Rotation Curves]{\label{fig:1b}\includegraphics[scale=.4]{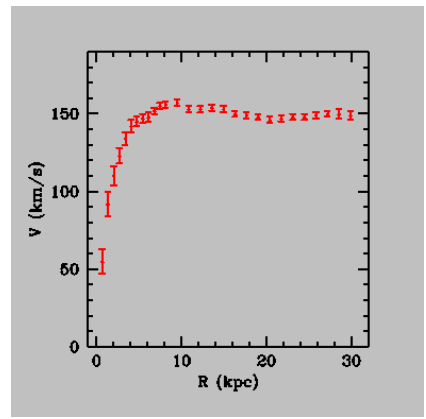}}
 \end{center}
 \caption{Galactic Rotation Curves} \label{fig:1}
\end{figure}

It appears that  there is five times more material in clusters of
galaxies than the galaxies and hot gas which we can see add up to.
That is most of the matter in clusters of galaxies is invisible and,
since these are the largest structures in the Universe held together
by gravity, many scientists then deduce that most of the matter in
the entire Universe is invisible. This invisible matter is called
'dark matter'. Over the years there has been a lot of research by
scientists attempting to discover exactly what this dark matter is,
how much there is, and what effect it may have on the future of the
 Universe as a whole.
Fritz Zwicky was the first to note in 1933 that the outlying
galaxies in the Coma cluster were moving much faster than mass
calculated for the visible galaxies  and this would indicate that
there is dark matter. Vera Rubin used galactic rotation curves to
deduce that there was dark matter in galaxies .These rotation curves
showed how average velocity of stars change with distance from
center of galaxy. The observed rotation curves differed from the
theoretical ones based on Keplerian orbits as can be seen in Figures
\ref{fig:1a} and \ref{fig:1b}.

\section{Varying G Dynamics}

we would first like to observe that even after all these decades
there is neither evidence for the hypothesized Dark matter, nor any
clue to its exact nature, if it exists. Let us see how varying $G$
dynamics removes the need for Dark Matter thus providing an
alternative explanation. Cosmologies with time varying $G$ have been
considered in the past, for example in the Brans-Dicke theory or in
the Dirac large number theory or  in the model of Hoyle
\cite{barrowparsons,narfpl,narburbridge,BG-Ha08,BG-csf03}. In the
case of the Dirac cosmology, the motivation was Dirac's observation
that the supposedly large number coincidences involving $N$, the
number of elementary particles in the universe had an underlying
message if it is recognized that
\begin{equation}
\sqrt{N} \propto T\label{3ea1}
\end{equation}
where $T$ is the age of the universe. Equation (\ref{3ea1}) leads to
a $G$ decreasing inversely with time in Dirac's hypothetical development.\\
The Brans-Dicke cosmology arose from the work of Jordan who was
motivated by Dirac's ideas to try and modify General Relativity
suitably. In this scheme the variation of $G$ could be obtained from
a scalar field $\phi$ which would satisfy a conservation law. This
scalar tensor gravity theory was further developed by Brans and
Dicke, in which $G$
was inversely proportional to the variable field $\phi$. (It may be mentioned
that more recently the ideas of Brans and Dicke have been further generalized.)\\
In the Hoyle-Narlikar steady state model, it was assumed that in the
Machian sense the inertia of a particle originates from the rest of
the matter present in the universe. This again has been shown to
 lead to a variable $G$. The above references give further details of
these various schemes and their shortcomings which have
lead to their falling out of favour.\\
Then there is  fluctuational cosmology in which particles are
fluctuationally created from a background dark energy, in an
inflationary type phase transition  and this leads to a scenario of
an accelerating universe with a small cosmological constant. This
1997 work \cite{BG-ijmpa98} was observationally confirmed a year
later due to the work of Perlmutter and others \cite{BG-cu}.
Moreover in  this cosmology, the various supposedly miraculous large
number coincidences as also the otherwise inexplicable Weinberg
formula which gives the mass of an elementary particle in terms of
the gravitational constant and the Hubble constant are also deduced
from the underlying theory rather than being ad hoc.To quote the
main result, the gravitational constant is given by
\begin{equation}
G = \frac{G_0}{T}\label{3ea2}
\end{equation}
where $T$ is time (the age of the universe) and $G_0$ is a constant.
Furthermore, other routine effects like the precession of the
perihelion of Mercury and the bending of light, and so on have also
been explained in this model. Moreover in this model, the
cosmological contant $\Lambda$ is given by $\Lambda \leq 0 (H^2)$
and shows an inverse dependence $1/T^2$ on time. We will  see that
there is
observational evidence for (\ref{3ea2}).\\
\indent With this background, we now give some tests for equation
(\ref{3ea2}).
\section{A test}
Let us first see the correct gravitational bending of light. In fact
in Newtonian theory too we obtain the bending of light, though the
amount is half that predicted by General
Relativity\cite{narlikarcos,denman,silverman,brill}. In the
Newtonian theory we can obtain the bending from the well known
orbital equations (Cf.\cite{gold66}),
\begin{equation}
\frac{1}{r} = \frac{GM}{L^2} (1+ecos\Theta)\label{3ey25}
\end{equation}
where $M$ is the mass of the central object, $L$ is the angular
momentum per unit mass, which in our case is $bc$, $b$ being the
impact parameter or minimum approach distance of light to the
object, and $e$ the eccentricity of the trajectory is given by
\begin{equation}
e^2 = 1+ \frac{c^2L^2}{G^2M^2}\label{3ey26}
\end{equation}
For the deflection of light $\alpha$, if we substitute $r = \pm
\infty$, and then use (\ref{3ey26}) we get
\begin{equation}
\alpha = \frac{2GM}{bc^2}\label{3ey27}
\end{equation}
This is half the General
Relativistic value.\\
We now observe that in this case we have,
\begin{equation}
G = G_o (1- \frac{t}{t_o})\label{3ey15}
\end{equation}
\begin{equation}
r = r_o \left(\frac{t_o}{t_o+t}\right)\label{3ey16}
\end{equation}

We now observe that the effect of time variation of $r$ is given by
equation (\ref{3ey16})(cf.ref.\cite{bgsnc115b}). Using this, the
well known equation for the trajectory is given by,
\begin{equation}
u'' + u = \frac{GM}{L^2} + u\frac{t}{t_0} + 0 \left (
\frac{t}{t_0}\right )^2\label{3ey28}
\end{equation}
where $u = \frac{1}{r}$ and primes denote differentiation with
respect to
$\Theta$.\\
The first term on the right hand side represents the Newtonian
contribution while the remaining terms are the contributions due to
(\ref{3ey16}). The solution of (\ref{3ey28}) is given by
\begin{equation}
u = \frac{GM}{L^2} \left[ 1 + ecos\left\{
\left(1-\frac{t}{2t_0}\right ) \Theta +
\omega\right\}\right]\label{3ey29}
\end{equation}
where $\omega$ is a constant of integration. Corresponding to
$-\infty < r < \infty$ in the Newtonian case we have in the present
case, $-t_0 < t < t_0$, where $t_0$ is large and infinite for
practical purposes. Accordingly the analogue of the reception of
light for the observer, viz., $r = + \infty$ in the Newtonian case
is obtained by taking $t = t_0$ in (\ref{3ey29}) which gives
\begin{equation}
u = \frac{GM}{L^2} + ecos \left(\frac{\Theta}{2} + \omega
\right)\label{3ey30}
\end{equation}
Comparison of (\ref{3ey30}) with the Newtonian solution obtained by
neglecting terms $\sim t/t_0$ in equations (\ref{3ey28}) and
(\ref{3ey29}) shows that the Newtonian $\Theta$ is replaced by
$\frac{\Theta}{2}$, whence the deflection obtained by equating the
left side of (\ref{3ey30}) to zero, is
\begin{equation}
cos \Theta \left(1-\frac{t}{2t_0}\right) = -\frac{1}{e}\label{3ey31}
\end{equation}
where $e$ is given by (\ref{3ey26}). The value of the deflection
from (\ref{3ey31}) is twice the Newtonian deflection given by
(\ref{3ey27}). That is the deflection $\alpha$ is now given not by
(\ref{3ey27}) but by the formula,
\begin{equation}
\alpha = \frac{4GM}{bc^2},\label{3ey32}
\end{equation}
The relation (\ref{3ey32}) is the correct observed value and is the
same as the General Relativistic formula which however is obtained
by a different route \cite{brill,berg,lass}.
\section{Galactic Rotation Curves and Dark Matter}
 We now come to the
problem of galactic rotational curves mentioned earlier
(cf.ref.\cite{narlikarcos}). We would expect, on the basis of
straightforward dynamics that the rotational velocities at the edges
of galaxies would fall off according to
\begin{equation}
v^2 \approx \frac{GM}{r}\label{3ey33}
\end{equation}
However  as seen in Section (1), it is found that the velocities
tend to a constant value,
\begin{equation}
v \sim 300km/sec\label{3ey34}
\end{equation}
as we approach the edges of the galaxies. This, as noted, has lead
to the postulation of the as yet undetected additional matter
alluded to, the so called dark matter.(However for an alternative
view point Cf.\cite{sivaramfpl93}). We observe that from
(\ref{3ey16}) it can be easily deduced that \cite{BG-cu,bgsedge}
\begin{equation}
a \equiv (\ddot{r}_{o} - \ddot{r}) \approx \frac{1}{t_o}
(t\ddot{r_o} + 2\dot r_o) \approx -2 \frac{r_o}{t^2_o}\label{3ey35}
\end{equation}
as we are considering infinitesimal intervals $t$ and nearly
circular orbits. Equation (\ref{3ey35}) shows
(Cf.ref\cite{bgsnc115b} also) that there is an anomalous inward
acceleration, as if there is an extra attractive force, or an
additional central mass.\\
So, we now have
\begin{equation}
\frac{GMm}{r^2} + \frac{2mr}{t^2_o} \approx
\frac{mv^2}{r}\label{3ey36}
\end{equation}
From (\ref{3ey36}) it follows that
\begin{equation}
v \approx \left(\frac{2r^2}{t^2_o} + \frac{GM}{r}\right)^{1/2}
\label{3ey37}
\end{equation}
From (\ref{3ey37}) it is easily seen that at distances within the
edge of a typical galaxy, that is $r < 10^{23}cms$ the equation
(\ref{3ey33}) holds but as we reach the edge and beyond, that is for
$r \geq 10^{24}cms$ we have $v \sim 10^7 cms$ per second, in
agreement with (\ref{3ey34}). In fact as can be seen from
(\ref{3ey37}), the first term in the square root has an extra
contribution (due to the varying $G$) which is roughly some three to
four times the second term, as if there is an extra mass, roughly
that much more. In fact the velocity at the edge of the galaxies as
calculated from equation (\ref{3ey35}) are tabulated in  the
following table, where the radius is in units of $10^{23}$ cm. The
table shows that equation (\ref{3ey35}) is in agreement with the
observed velocity given in equation (\ref{3ey32}).

\[ \left|\begin{array} {cc} \vspace{.2cm}
{\it Velocity} & {\it Radius}\\
\hline
 8.124038527  \times 10^7 & 0.01 \\
 3.316635644 \times 10^7 & 0.06\\
 2.449539140\times 10^7 & 0.11\\
 2.031135642 \times 10^7 & 0.16\\
 1.773059260 \times 10^7 & 0.21\\
 1.593679245 \times 10^7 &  0.26\\
 1.459778838 \times 10^7 & 0.31 \\
 1.354963222 \times 10^7 & 0.36\\
 1.270085862 \times 10^7 & 0.41\\
 1.199589350 \times 10^7 & 0.46\\
 1.139877031 \times 10^7 & 0.51\\
 1.088505134 \times 10^7 & 0.56 \\
 1.043747676 \times 10^7 &  0.61\\
 1.004346553 \times 10^7 & 0.66\\
 9.693603379 \times 10^6 & 0.71\\
 9.38068788 \times 10^6 & 0.76\\
 9.09910333 \times 10^6 & 0.81\\
 8.84439856 \times 10^6 & 0.86\\
 8.612994399\times 10^6 & 0.91\\
 8.401975958 \times 10^6 & 0.96\\
 8.208942359\times 10^6 &1.01\\
 8.03189584 \times 10^6 & 1.06\\
 7.869158751 \times 10^6 & 1.11\\
 7.719310314 \times 10^6 & 1.16\\
 7.581138077 \times 10^6 & 1.21\\
 7.453599961 \times 10^6 & 1.26\\
 7.335794393 \times 10^6 & 1.31\\
 7.226936540 \times 10^6 & 1.36\\
 7.126339217 \times 10^6 & 1.41\\
 7.033397433 \times 10^6 & 1.46\\
 6.947575783 \times 10^6 &1.51\\
 6.868398088 \times 10^6 & 1.56\\
 6.795438824 \times 10^6 & 1.61\\
 6.728315996\times 10^6 & 1.66\\
 6.666685175 \times 10^6 & 1.71\\
 6.610234489 \times 10^6 & 1.76\\
 6.558680385 \times 10^6 &1.81\\
 6.511764044 \times 10^6 &1.86\\
 6.469248319 \times 10^6 &1.91\\
 6.430915128 \times 10^6 & 1.96.
\end{array}\right|\]
 Thus the time variation of $G$ explains observation
without invoking dark matter.It may be added that this also explains
the latest studies by Metz,Kroupa and others of the satellite
galaxies of the Milky Way galaxy , which also throw up the faster
than expected rotational velocities, ruling out however, in this
case dark matter, in addition.
\section{Remarks}
 There could be other explanations,
too. One of the authors and A.D. Popova have argued that if the
three dimensionality of space asymptotically falls off, then the
above can be explained \cite{bgspopovadeds}.\\ Yet another
prescription was given by Milgrom \cite{milgrom} who approached the
problem by modifying Newtonian dynamics at large distances. It must
be mentioned that this approach is purely
phenomenological.\\
The idea was that perhaps standard Newtonian dynamics works at the
scale of the solar system but at galactic scales involving much
larger distances perhaps the situation is different. However a
simple modification of the distance dependence in the gravitation
law, as pointed by Milgrom would not do, even if it produced the
asymptotically flat rotation curves of galaxies. Such a law would
predict the wrong form of the mass velocity relation. So Milgrom
suggested the following modification to Newtonian dynamics: A test
particle at a distance $r$ from a large mass $M$ is subject to the
acceleration $a$ given by
\begin{equation}
a^2/a_0 = MGr^{-2},\label{3em1}
\end{equation}
where $a_0$ is an acceleration such that standard Newtonian dynamics
is a good approximation only for accelerations much larger than
$a_0$. The above equation however would be true when $a$ is much
less than $a_0$. Both the statements in (\ref{3em1}) can be combined
in the heuristic relation
\begin{equation}
\mu (a/a_0) a = MGr^{-2}\label{3em2}
\end{equation}
In (\ref{3em2}) $\mu(x) \approx 1$ when $x >> 1, \, \mbox{and}\,
\mu(x) \approx x$ when $x << 1$. It must be stressed that
(\ref{3em1}) or (\ref{3em2}) are not deduced from any theory, but
rather are an ad hoc prescription to explain observations.
Interestingly it must be mentioned that most of the implications of
Modified Newtonian Dynamics or MOND do not
depend strongly on the exact form of $\mu$.\\
It can then be shown that the problem of galactic velocities is now
solved \cite{mil1,mil2,mil3,mil4,mil5}. \\ Finally it maybe
mentioned that the above varying $G$ dynamics explains the puzzling
anomalous acceleration of the Pioneer spacecrafts of the order of
$10^-7$ to $10^-8$ $cm/sec^2$ observed by J.D.Anderson of the JPL
Pasadena, and co-workers \cite{BG-tu08}.

\bibliographystyle{amsplain}
\bibliography{}
\end{document}